\documentclass[epjCONF, onecolumn]{svjour}
\usepackage{graphics}
\usepackage[varg]{txfonts} 
\usepackage[latin1]{inputenc}
\session-title{Conference Title, to be filled}
\begin{document}
\title{Schwarzschild models of the Sculptor dSph galaxy}
\author{Maarten A. Breddels\inst{1}\fnmsep\thanks{\email{breddels@astro.rug.nl}} \and Amina Helmi\inst{1} \and R.C.E. van den Bosch\inst{2} G. van de Ven\inst{2} \and G. Battaglia\inst{3}}
\institute{Kapteyn Astronomical Institute, University of Groningen, P.O. Box 800, 9700 AV Groningen, the Netherlands \and Max Planck Institute for Astronomy, K\"onigstuhl 17, 69117
Heidelberg, Germany \and \parbox[t]{\textwidth}{ European Organisation for Astronomical Research in the Southern
Hemisphere, K. Schwarzschild-Str. 2, 85748 Garching bei M\"unchen,
Germany}}
\abstract{We have developed a spherically symmetric dynamical model of a dwarf spheroidal galaxy using the Schwarzschild method. This type of modelling yields constraints both on the total mass distribution (e.g. enclosed mass and scale radius) as well as on the orbital structure of the system modelled (e.g. velocity anisotropy). Therefore not only can we derive the dark matter content of these systems, but also explore possible formation scenarios. Here we present preliminary results for the Sculptor dSph.
We find that the mass of Sculptor within 1kpc is $8.5\times10^{7\pm0.05}$M$_{\odot}$, its anisotropy profile is tangentially biased and slightly more isotropic near the center. For an NFW profile, the preferred concentration ($\sim$15) is compatible with cosmological models. Very cuspy density profiles (steeper than NFW) are strongly disfavoured for Sculptor.
} 

\maketitle

\section{Introduction}

The existence of dark matter has been invoked to explain discrepancies in the observed kinematics of (systems of) galaxies.
Dwarf spheroidal galaxies appear to be one of the most dark matter dominated galaxies, with total dynamical mass to stellar light ratios in the order of 100-1000 $M_\odot/L_\odot$. The nearby dSph galaxies have the additional advantage that individual stars can be resolved, and their red giant branch stars are bright enough to measure line-of-sight velocities with errors of a few km s$^{-1}$. Their high dynamical mass-to-light ratios makes these systems ideal to study dark matter halos, especially their internal structure and to constrain their inner density profiles. To this end we study the Sculptor dwarf spheroidal galaxy using orbit based dynamical models, also known as Schwarzschild models.

\section{Method and data}

\begin{figure}
\centering{
\resizebox{0.35\columnwidth}{!}{ \includegraphics{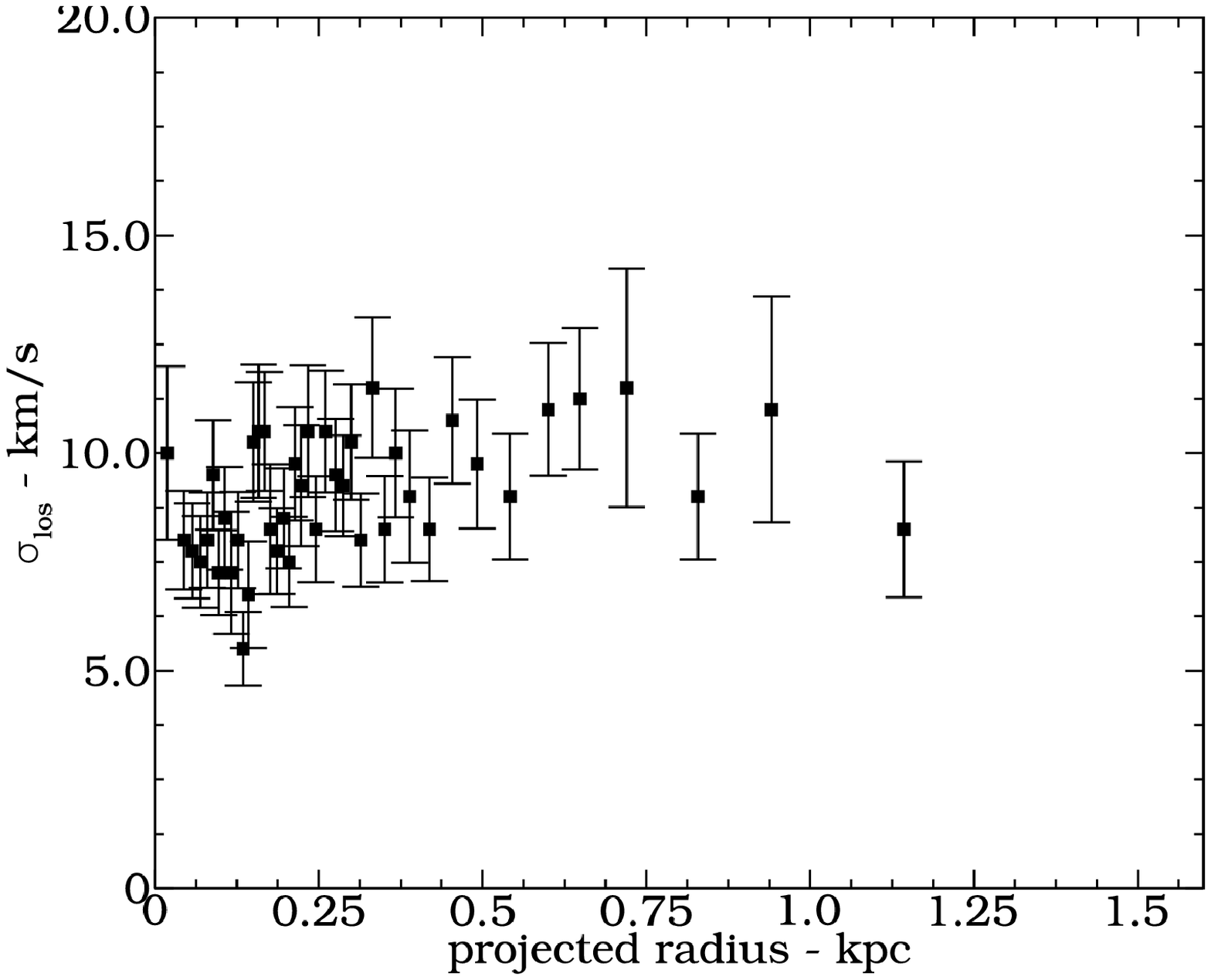}   }
\resizebox{0.32\columnwidth}{!}{ \includegraphics{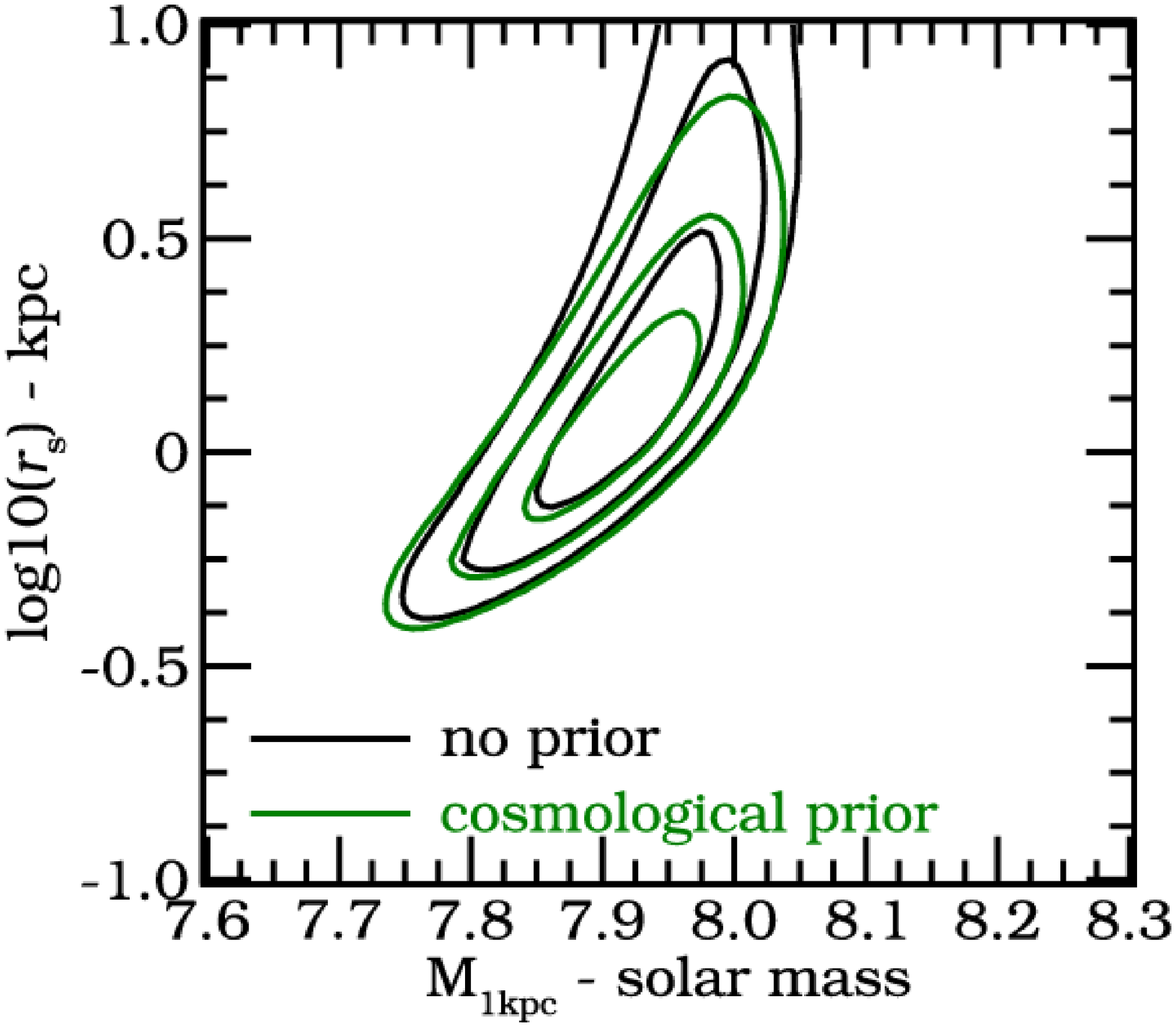} }}
\caption{Left: Velocity dispersion profile for Sculptor. Right: Joint probability density functions for the enclosed mass within 1 kpc and the NFW scale parameter ($r_s$) for Sculptor. }
\label{fig:1}       
\end{figure}

In our Schwarzschild models, we assume a specific gravitational potential in which we integrate test particle orbits. These orbits are drawn from a dense grid in the space of integrals of motion (in our case energy and  angular momentum). Using quadratic programming we find the set of weights such that the superposition of the weighted orbits matches the data. In our case we fit the light distribution, which we assume to be a Plummer sphere, and the velocity dispersion profile. For Sculptor we use the 2304 line of sight velocities from the datasets by \cite{Bat} and \cite{Walker}. In the left panel of Fig. \ref{fig:1} we show the velocity dispersion profile we fit.

\begin{figure}
\centering{
\resizebox{0.33\columnwidth}{!}{ \includegraphics{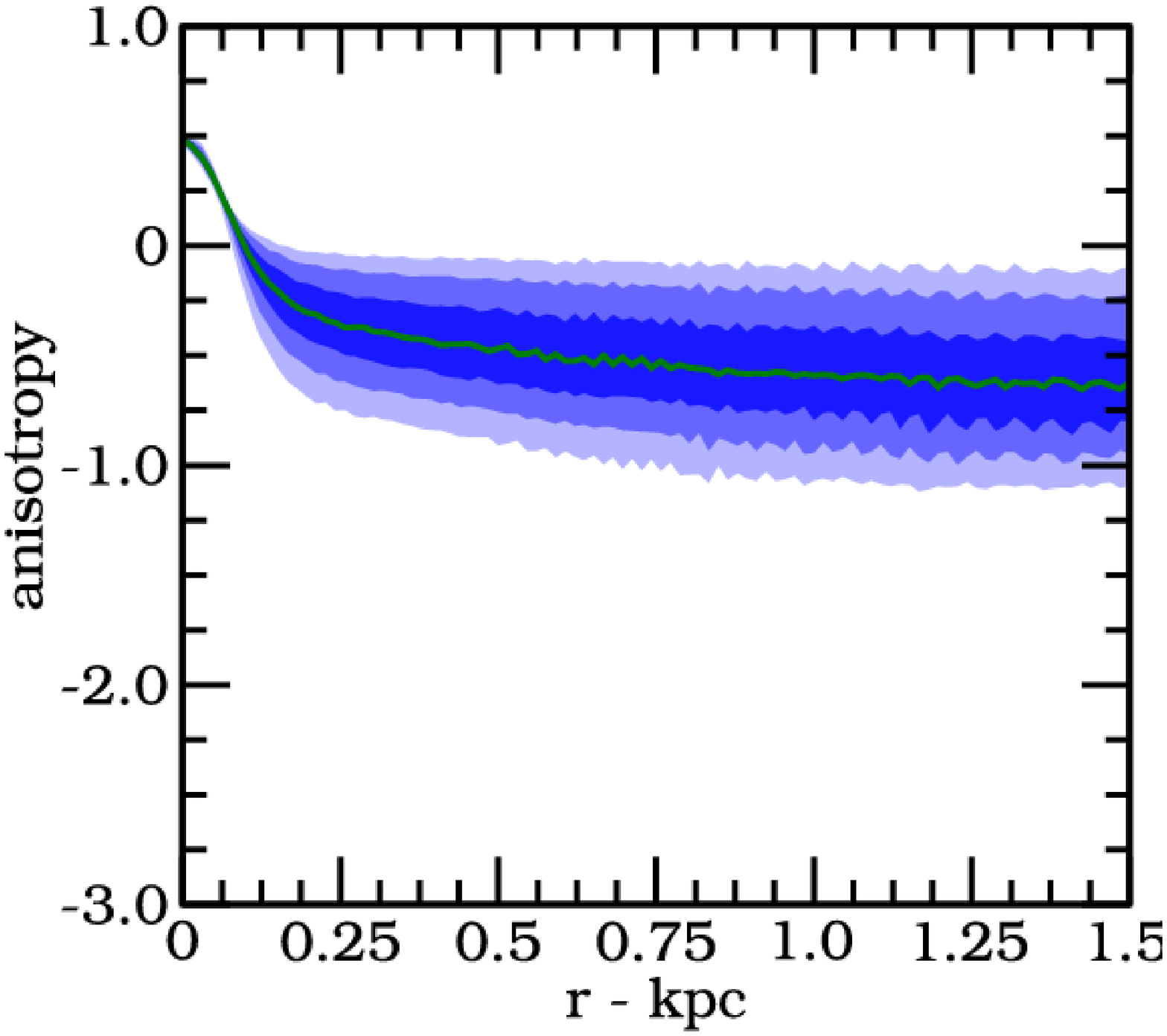}   }
\resizebox{0.308\columnwidth}{!}{ \includegraphics{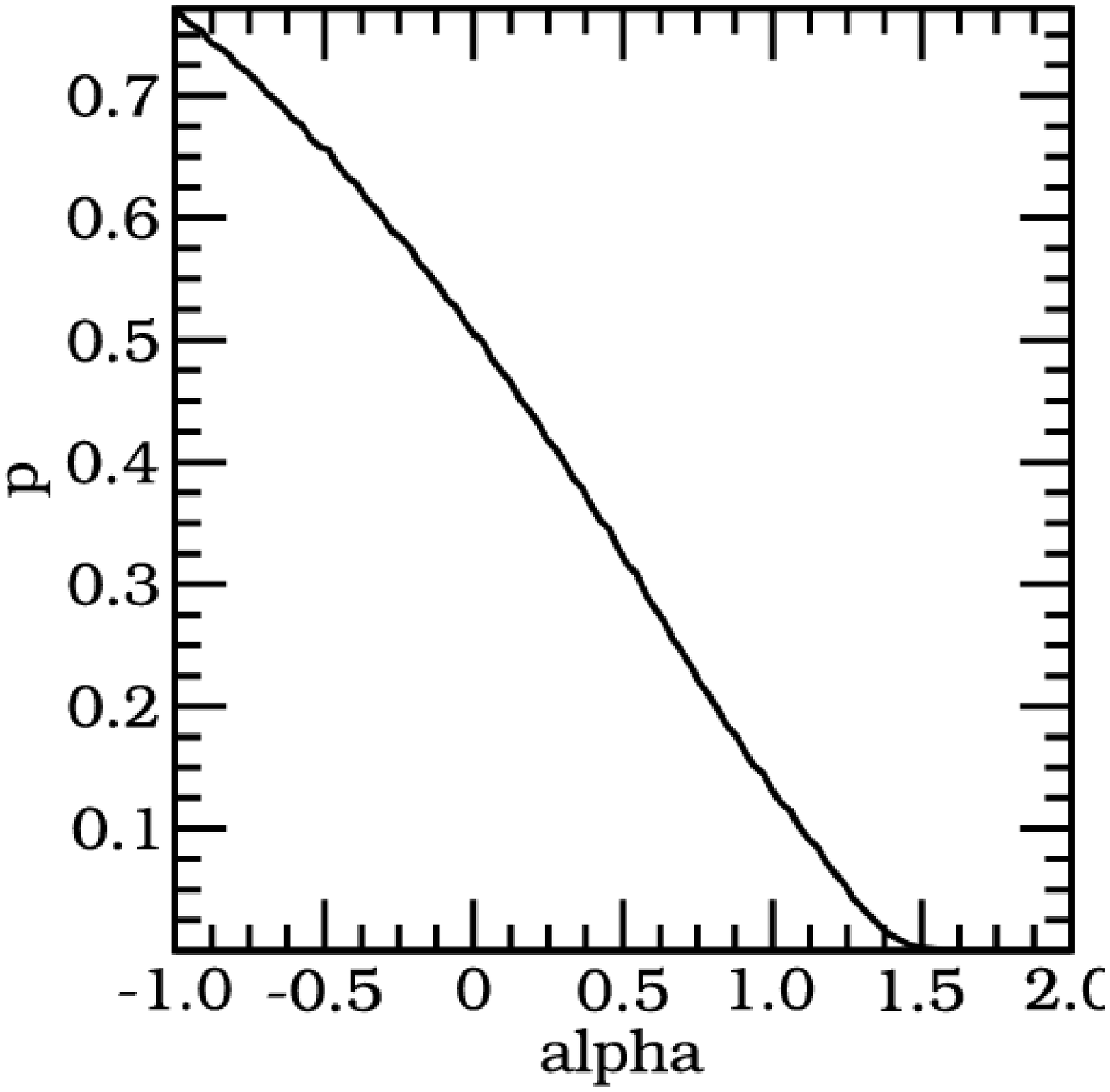} }
}
\caption{ Left: Resulting anisotropy profiles for Sculptor. Right: Probability density function of the inner slope of the dark matter density profile for Sculptor.}
\label{fig:2}       
\end{figure}

\section{Results}

In the right panel of Fig. \ref{fig:1} we show the constraints on the mass and scale radius ($r_s$) in the case an assumed NFW \cite{NFW} profile for Sculptor. The black contours show the 1, 2 and 3$\sigma$ confidence interval. It shows that although M(r$<$1kpc) is well constrained, the scale parameter ($r_s$) is less so. The concentration ($r_{200}/r_s \approx 15$) implied is compatible with cosmological models.  The green contour includes a cosmological prior (the mass-concentration relation from \cite{Maccio}), which gives a slightly better constraint on $r_s$. Since the method finds the distribution function, we can also reconstruct the orbital structure. In the left panel of Fig \ref{fig:2} we show the anisotropy profile in the left panel of Fig. \ref{fig:2}. The green line shows the median value, and the blue regions the 1,2,3 $\sigma$ confidence intervals. The velocity ellipsoid of Sculptor is more isotropic in the center compared to the outer regions. We also change the inner slope ($\alpha$) of the DM density distribution $\rho(r)=\rho_0(r/r_s)^{-\alpha}(1+r/r_s)^{-2}$ (for NFW, $\alpha=1$). The probability density function of $\alpha$ is shown in the right panel of Fig. \ref{fig:2}. Although the slope is not well constrained, very cuspy profiles ($\alpha>1.5$) seem to be ruled out for  Sculptor. More results can be found in \cite{Breddels}.

\end{document}